\documentclass[structabstract]{aa}

\usepackage{txfonts}
\usepackage{graphicx}
\usepackage{natbib}
\bibpunct[; ]{(}{)}{,}{a}{}{;} 

\begin{document}

\title{Broad-band Spectroscopy of the Ongoing Large Eruption of the Luminous Blue Variable R71\thanks{Based on observations collected at ESO's Very Large Telescope under Prog-IDs: 69.D-0390(D) and 289.D-5040(A) and at the MPG/ESO 2.2-m Telescope under Prog-IDs: 076.D-0609(A), 078.D-0790(B), 086.D-0997(A), and 087.D-0946(A).}
\fnmsep \thanks{This work was co-funded under the Marie Curie Actions of the European Commission (FP7-COFUND).}\\ 
} 

%\subtitle{JHK photometry of Eta Carinae with IRSF/SIRIUS} 

\author{A. Mehner\inst{1}
  \and D. Baade\inst{2} 
     \and T. Rivinius\inst{1}
     \and D. J. Lennon\inst{3}
       \and C. Martayan\inst{1} 
     \and O. Stahl\inst{4}
  \and S. \v{S}tefl\inst{5} } 

\offprints{A. Mehner, \email{amehner@eso.org}}

\institute{ 
  \and 
  \and } 

\institute{ESO -- European Organisation for Astronomical Research in the Southern Hemisphere, Alonso de Cordova 3107, Vitacura, Santiago de Chile, Chile 
  \and ESO -- European Organisation for Astronomical Research in the Southern Hemisphere, Karl-Schwarzschild-Stra{\ss}e 2,  85748 Garching, Germany
  \and ESAC -- European Space Astronomy Centre, Camino bajo del Castillo, Urbanizacion Villafranca del Castillo, Villanueva de la Ca\~nada, 28692 Madrid, Spain
  \and Zentrum f\"ur Astronomie der Universit\"at Heidelberg, Landessternwarte,
K\"onigstuhl 12, 69117 Heidelberg, Germany
    \and ESO/ALMA -- The European Organisation for Astronomical Research in the Southern
Hemisphere/The Atacama Large Millimeter/Submillimeter Array, Alonso de Cordova 3107, Vitacura, Santiago de Chile, Chile}   

%\date{Received 2 November 1992 / Accepted 7 January 1993}

\abstract {} {The Luminous Blue Variable (LBV) R71 is currently undergoing an eruption, which differs photometrically and spectroscopically from its last outburst in the 1970s. Valuable information on the physics of LBV eruptions can be gained by analyzing the spectral evolution during this eruption and by comparing R71's present appearance to its previous outburst and its quiescent state.} {An ongoing monitoring program with VLT/X-shooter will secure key spectral data ranging from visual to near-infrared wavelengths. Here we present the first spectra obtained in 2012 and compare them to archival VLT/UVES and MPG/ESO-2.2m/FEROS spectra from 2002 to 2011. The discussed data include pre-eruption spectra in 2002 and 2005, a spectrum of the transitionary phase between quiescent and eruptive state in 2007, and spectra of the eruption in 2011--2012. Information on R71's 1970s outburst is taken from the literature.} {The 2011--2012 spectra are dominated by strong neutral and singly ionized metal absorption lines likely formed in a large ``pseudo-photosphere.'' We find an unusually low apparent temperature of R71 of only $T_{\textnormal{\scriptsize{eff,2012}}} \sim$~6\,650~K; the star resembles a late F supergiant.  R71's visual lightcurve had a maximum in 2012 with $m_{\textnormal{\scriptsize{V,2012}}}\sim8.7$~mag. Given the uncertainty in the extinction towards R71, this corresponds to a bolometric luminosity of $M_{\textnormal{\scriptsize{bol,2012}}}\sim-9.8$~mag to $-10.3$~mag. R71's 2011--2012 spectra do not show \ion{H}{I} and \ion{Fe}{II} P Cyg profiles, which were present during its last outburst in the 1970s and which are normally observed during LBV outbursts.  Low-excitation forbidden emission lines and \ion{Fe}{I} P Cyg-like profiles from a slowly expanding nebula became apparent in late 2012. These lines originate likely in the rarefied region above the pseudo-photosphere up to 13~AU from the star.} {The rise in R71's visual magnitude and the low apparent temperature of its pseudo-photosphere during the current eruption are unprecedented for this star.  R71 most likely increased its bolometric luminosity by $\Delta M_{\textnormal{\scriptsize{bol}}} = 0.4$--$1.3$~mag compared to its quiescent state. The very low temperature of its pseudo-photosphere implies a very high-mass loss rate on the order of  $\dot M_{\textnormal{\scriptsize{R71,2012}}} \sim$~5$\times$10$^{-4}$~$M_{\odot}$~yr$^{-1}$ compared to  $\dot M_{\textnormal{\scriptsize{quiescence}}} \sim$~3$\times$10$^{-7}$~$M_{\odot}$~yr$^{-1}$. The apparent radius increased by a factor of 5 to about 500~$R_{\odot}$. No fast-moving material indicative of an explosion is observed. The changes in R71's photometry and spectrum are thus likely consequences of a tremendously increased wind density, which led to the formation of a pseudo-photosphere. 
%This gives clues about the instability process, since mechanisms that involve ejection of material at velocities higher than the normal stellar wind velocity can be excluded.
} 


\maketitle

\section{Introduction}

The LBV R71 (= HDE 269006) in the Large Magellanic Cloud (LMC) is currently undergoing an eruption that started in 2005 \citep{2009IAUC.9082....1G}. In 2012 the star had reached unprecedented visual brightness, accompanied by remarkable variations in its optical spectrum \citep{2012CBET.3192....1G}. R71's current eruption is characterized by more extreme parameters than its outburst in the 1970s and may provide us with important insights into the LBV phenomenon. 

LBVs, also known as S Doradus variables, are evolved massive stars that exhibit a particular type of instability which is not yet understood (\citealt{1984IAUS..105..233C,1997ASPC..120..387C,1994PASP..106.1025H,1997ASPC..120.....N}, and references therein). They represent a brief but critical phase in massive star evolution because several solar masses can be expelled during this stage. Most of the fundamental questions about the physical cause of their instability are still unsolved and they may fall into different categories, i.e., classical LBV outbursts and giant eruptions. 
Outbursts with visual magnitude variations of 1--2~mag and constant bolometric luminosity are commonly referred to as classical LBV outbursts. During giant eruptions the visual magnitude increases by more than 2~mag and the bolometric luminosity likely increases.  Historical examples of giant eruptions are P Cygni in the 17th century (e.g., \citealt{1988IrAJ...18..163D,1992A&A...257..153L}), $\eta$ Car in the 1840s (see the reviews by \citealt{1997ARA&A..35....1D,2012ASSL..384.....D} and references therein), V12 in NGC 2403 in the 1950s (e.g., \citealt{1968ApJ...151..825T,2001PASP..113..692S}), and SN1961V in NGC 1058  (e.g., \citealt{1989ApJ...342..908G,2012ApJ...746..179V}; cf.\ \citealt{2011ApJ...737...76K}).   The most promising explanation for the LBV instability mechanism involve radiation pressure instabilities, but also turbulent pressure instabilities, vibrations and dynamical instabilities, and binarity cannot be entirely ruled out (\citealt{1994PASP..106.1025H} and references therein). It is unknown what the relative roles of classical LBV outbursts and giant eruptions are and which of the two events are more important in terms of total mass lost. It is also not established if giant eruptions result from unusual circumstances and due to a different instability mechanism or if they are only extreme cases of classical LBV outbursts.

LBVs may be critical to our understanding of massive star evolution. 
Their instability could be responsible for the empirically found upper luminosity boundary (also known as Humphreys-Davidson limit) above and to the right of which no supergiants are found \citep{1979ApJ...232..409H,1984Sci...223..243H}.  The commonly accepted idea is that very high-mass loss prevents stars above $40$--$50~M_\odot$ from becoming red supergiants (RSGs) but instead drives them towards higher temperatures. LBVs are located in the region of concern in the HR-diagram and their instability leads to high-mass loss episodes. 
The realization that winds of O and Wolf-Rayet (WR) stars are clumpy/porous \citep{2006ApJ...637.1025F,2008A&ARv..16..209P} has reduced their inferred mass loss rates by a factor of $\sim$10. This further increased the interest in the LBV phase, since high-mass loss $\emph{episodes}$ may be required for massive stars to evolve directly to the WR stage \citep{2006ApJ...645L..45S}.

LBVs are generally considered to be stars in transition to the WR stage \citep{1983A&A...120..113M,1994PASP..106.1025H,1994A&A...290..819L}, but observations of LBV-like eruptions immediately prior to supernova (SN) explosions and also recent stellar evolution modeling may lead to a revision of this interpretation. 
For several exceedingly luminous SNe it has been suggested that shock interaction with dense and opaque circumstellar material could account for the excess luminosity \citep{2007ApJ...666.1116S,2008ApJ...686..467S}. The progenitor stars likely expelled extensive circumstellar ejecta during LBV outbursts or giant eruptions. \citet{2012MNRAS.423L..92Q} proposed an alternative explanation for large stellar mass-loss rates just prior to core collapse. During core Ne and O fusion in massive stars the fusion luminosity exceeds the star's Eddington luminosity, which drives convection and in turn can excite internal gravity waves. The gravity waves convert into sound waves as they propagate towards the stellar surface and the subsequent dissipation of the sound waves can unbind several solar masses of the stellar envelope.

In a few cases LBV-like eruptions have been observed prior to a SN explosion; SN2006jc \citep{2007ApJ...657L.105F,2007Natur.447..829P}, SN2005gl \citep{2007ApJ...656..372G,2009Natur.458..865G}, and SN2009ip \citep{2012arXiv1209.6320M} but compare  \citet{2011MNRAS.412.1639D} who argued that only in the case of SN2005gl there is substantial evidence for an LBV progenitor. 
In the recent case of SN2009ip, \citet{2010AJ....139.1451S} identified the progenitor as a likely LBV based on its photometry but its multiple pre-SN outbursts were most likely due to pulsational pair instability, outbursts associated with late stages of nuclear burning, or close periastron passages of a companion star rather than wind-driven LBV outbursts because their time scales were much shorter than for most other LBV outbursts \citep{2012arXiv1209.6320M}. \citet{2013ApJ...764L...6S} proposed that SN2009ip's pre-SN outbursts were due to binary interaction and that its final explosion was a mergerburst and not a true SN. 
\citet{2013Natur.494...65O} claimed that a mass loss event of the progenitor of SN2010mc is consistent with a wave-driven pulsation model and is causally connected to its SN explosion only 40 days after.

The link between SNe and LBVs had not been supported by stellar evolutionary models until  \citet{2013A&A...550L...7G} reported on their theoretical finding that single rotating stars with initial mass in the range of 20--25$~M_{\odot}$ have spectra similar to LBVs before exploding as SNe. This result sets the theoretical ground for low-luminosity LBVs, such as possibly R71 (but see Section \ref{discussion}), to be the endpoints of stellar evolution and may also strengthen the identification of the progenitor of SN1987A as an LBV \citep{2007AJ....133.1034S}.

Important clues towards an understanding of the LBV phenomenon can be gained from their variabilities that occur on different magnitude and time scales (\citealt{1994PASP..106.1025H} and references therein). R71 is excellent for a case study.
It has been well-observed over the last decades and its location in the LMC is fortunate because its distance and reddening are well constrained.
During its quiescent phase R71 has a hot supergiant spectrum with strong \ion{H}{I}, \ion{He}{I}, \ion{Fe}{II}, and [\ion{Fe}{II}] lines and P Cyg profiles. It was classified as a B2.5~Ieq star with $m_{\textnormal{\scriptsize{V}}}\sim10.9$~mag \citep{1960MNRAS.121..337F}. It has a typical supergiant mass loss rate of $\sim$3$\times$10$^{-7}$ $M_\odot$ yr$^{-1}$  \citep{1981A&A...103...94W}.
%Microvariations with $\Delta m_{\textnormal{\scriptsize{V}}}\sim0.2$ mag occur on time scales of weeks to months. 
The star shows microvariations of $\Delta m_{\textnormal{\scriptsize{V}}}\sim0.1$~mag on time scales of 14--100~days \citep{1985A&A...153..163V,1988A&AS...74..453V,1997A&AS..124..517V,1998A&A...335..605L}.
%Classical LBV outbursts with $\Delta m_{\textnormal{\scriptsize{V}}}\sim0.5-2.0$ mag occur on time scales of years to a decade. The variations in the visual are due to the formation of a pseudo-photosphere while the bolometric luminosity stays constant.  
R71 had a classical LBV outburst in 1970--1977 when it reached  $m_{\textnormal{\scriptsize{V}}}\sim9.9$~mag \citep{1974MNRAS.168..221T,1975A&A....41..471W,1979A&AS...38..151V,1981A&A...103...94W,1982A&A...112...61V}. During the outburst its mass loss rate increased by a factor of about 100 to  $\sim$5$\times$10$^{-5}$ $M_\odot$ yr$^{-1}$  (\citealt{1981A&A...103...94W}, cf.\ \citealt{2002A&A...393..543V} who found that R71's mass loss rate increased only by a factor of 3--5).

R71's current eruption started in 2005. In 2012 the star had reached an unprecedented visual magnitude of $m_{\textnormal{\scriptsize{V}}}\sim8.7$~mag and is currently the visually brightest star in the LMC (http://www.aavso.org; cf.\ \citealt{1960MNRAS.121..337F}). At the end of 2012 a decline in R71's visual lightcurve commenced while the R-band lightcurve stayed bright (http://www.aavso.org). \citet{2009IAUC.9082....1G,2012CBET.3192....1G} found that the spectrum in 2008 February resembled an early-A supergiant, in 2009 August an extreme early-F hypergiant, and in 2012 that of an  early-G supergiant. 

In this paper we discuss R71's current eruption in relation to its previous outburst in the 1970s and to its quiescent state. Because of the rarity of LBV outbursts and their potential importance in massive star evolution this large-amplitude event calls for special attention. In Section \ref{obs} we describe the observations. In Section \ref{results} we analyze the changes of selected spectral features in detail. In Section \ref{discussion} we discuss the results and consider R71's classification as an underluminous or classical LBV. In Section \ref{conclusion} we summarize our conclusions.

\section{Observations and Data Analysis}
\label{obs}

\begin{table*}
\caption{Journal of observations.} % title of Table
\label{table:journal} % is used to refer this table in the text
\centering % used for centering table
\begin{tabular}{lllll} % centered columns (4 columns)
\hline\hline % inserts double horizontal lines
Date & MJD  & Instrument   & Wavelength Range & Total Exposure Time   \\ % table heading
 & (days)  & & (\AA) & (s)  \\ % table heading
\hline % inserts single horizontal line
2002-07-23	&	52478.4	&	UVES    	&	3\,260--4\,450 	&	 	360 \\
2002-07-23	&	52478.4	&	UVES    	&	4\,580--6\,680	&	3x110 \\
2002-08-03	&	52489.4	&	UVES    	&	3\,260--4\,450  &	360 \\
2002-08-03	&	52489.4	&	UVES    	&	4\,580--6\,680	&	3x110 \\
2002-08-12	&	52498.4	&	UVES    	&	3\,260--4\,450	&	360 \\
2002-08-12	&	52498.4	&	UVES    	&	4\,580--6\,680	&	3x110 \\
2002-09-04    &	52521.4	&	UVES    	&	3\,260--4\,450	&	360 \\
2002-09-04	&	52521.4	&	UVES    	&	4\,580--6\,680	&	3x110 \\
2002-09-10	&	52527.4	&	UVES    	&	3\,260--4\,450	&	360 \\
2002-09-10	&	52527.4	&	UVES    	&	4\,580--6\,680	&	3x110 \\
2002-09-18	&	52535.3	&	UVES    	&	3\,260--4\,450	&	360 \\
2002-09-18    &	52535.3	&	UVES    	&	4\,580--6\,680	&	3x110 \\
2002-09-22	&	52539.3	&	UVES    	&  3\,260--4\,450	&	360 \\
2002-09-22	&	52539.3	&	UVES    	&	4\,580--6\,680	&	3x110 \\
2002-09-30	&	52547.4	&	UVES    	&	3\,260--4\,450	&	360 \\
2002-09-30	&	52547.4	&	UVES    	&	4\,580--6\,680	&	3x110 \\
2002-10-06	&	52553.4	&	UVES    	&	3\,260--4\,450	&	360 \\
2002-10-06	&	52553.4	&	UVES    	&	4\,580--6\,680	&	3x110 \\
2005-12-12	&	53716.2	&	FEROS   	&	3\,500--9\,200	&	2x450	\\
2007-02-22	&	54153.2	&	FEROS   	&	3\,500--9\,200	&	3x500	\\
2011-03-20	& 55641.0	&	FEROS   	&	3\,500--9\,200	&	1800	\\
2011-05-15	&	55697.0	&	FEROS   	&	3\,500--9\,200	&	1800	\\
2012-08-26	&	56165.3	&	X-shooter	&	10\,240--24\,800	&	3x60\tablefootmark{a}, 2x10\tablefootmark{b}	\\
2012-08-26 	&	56165.3	&	X-shooter	&	3\,000--5\,595	&	3x120\tablefootmark{a}, 2x20\tablefootmark{b}	\\
2012-08-26 	&	56165.3	&	X-shooter	&	5\,595--10\,240	&	3x120\tablefootmark{a}, 2x20\tablefootmark{b}	\\
2012-09-04	&	56174.3	&	X-shooter	&	10\,240--24\,800	&	3x60\tablefootmark{a}, 2x10\tablefootmark{b}	\\
2012-09-04 	&	56174.3	&	X-shooter	&	3\,000--5\,595	&	3x120\tablefootmark{a}, 2x20\tablefootmark{b}	\\
2012-09-04 	&	56174.3	&	X-shooter	&	5\,595--10\,240	&	3x120\tablefootmark{a}, 2x20\tablefootmark{b}	\\
2012-11-01	&	56232.2	&	X-shooter	&	10\,240--24\,800	&	3x60\tablefootmark{a}, 2x10\tablefootmark{b}	\\
2012-11-01 	&	56232.2	&	X-shooter	&	3\,000--5\,595	&	3x120\tablefootmark{a}, 2x20\tablefootmark{b}	\\
2012-11-01 	&	56232.2	&	X-shooter	&	5\,595--10\,240	&	3x120\tablefootmark{a}, 2x20\tablefootmark{b}	\\
\hline %inserts single line
\end{tabular}
\tablefoot{
\tablefoottext{a}{Slit width: 0\farcs5/0\farcs4/0\farcs4 for the UVB/VIS/NIR arms.}
\tablefoottext{b}{Slit width: 5{\arcsec}/5{\arcsec}/5{\arcsec} for the UVB/VIS/NIR arms.}
}
\end{table*}

 \begin{figure*}
\centering
\resizebox{\hsize}{!}{\includegraphics{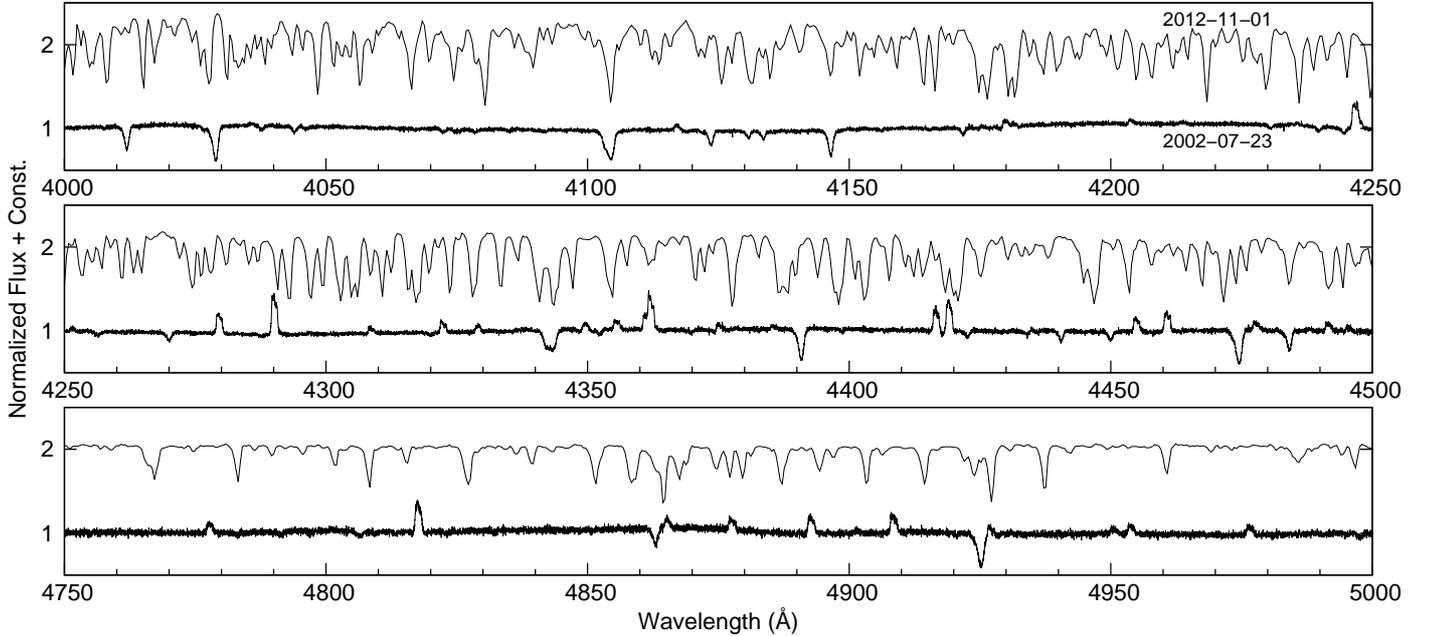}}
     \caption{Spectral variations between R71's quiescent state in 2002 July and its eruptive state in 2012 November.  The 2002 UVES spectrum shows strong [\ion{Fe}{II}] emission and \ion{Fe}{II} P Cyg profiles (e.g., at $\lambda4924$ \AA). H$\beta$ has a P Cyg  profile, while higher Balmer lines appear only in absorption. The 2012 X-shooter spectrum is dominated by strong metal absorption lines. (The continuum is normalized to unity. The wavelength region $\lambda\lambda4500$--$4750$ \AA\ is not displayed because of a gap in wavelength between the UVES blue and red settings.)}
     \label{figure:overview}
\end{figure*}

In 2012 August, September, and November we obtained the first X-shooter spectra of R71 of a long-term monitoring program at the Very Large Telescope (VLT). X-shooter is a medium-resolution echelle spectrograph that simultaneously observes with 3 arms covering the wavelength region from $3\,000$--$24\,800$~\AA\ \citep{2011A&A...536A.105V}. 
Spectra were obtained with the narrowest available slits of 0\farcs5 in the UVB arm, 0\farcs4 in the VIS arm, and 0\farcs4 in the NIR arm yielding spectral resolving powers of $R\sim10\,000$--$18\,000$. In addition, spectra were obtained  with the 5\arcsec\ slits in all 3 arms to achieve an absolute flux calibration but also to investigate the spectral energy distribution.
 
We retrieved archival spectra of R71 obtained with  VLT/UVES in 2002 and with MPG/ESO-2.2m/FEROS in 2005, 2007, and 2011. The UVES spectra in 2002 were obtained during R71's quiescent state. They cover the wavelength region from $3\,000$--$6\,800$~\AA\ with spectral resolving power of $R\sim40\,000$. The FEROS spectra cover the wavelength region from $3\,500$--$9\,200$~\AA\ with spectral resolving power of $R\sim48\,000$ and were obtained throughout the current outburst starting at its onset in 2005.

A journal of observations is listed in Table \ref{table:journal}. Each data set was reduced with the corresponding ESO pipeline (X-shooter pipeline version 1.5.0, UVES pipeline version 5.0.17, FEROS pipeline version 1.60). The spectra were corrected for the instrument response and atmospheric extinction but not for interstellar extinction and telluric lines. 

We used the 2002 UVES spectra to determine R71's systemic velocity. During the quiescent state the permitted lines show P Cyg profiles and we therefore  used the [\ion{Fe}{II}] emission lines at $\lambda\lambda$4287,4414,4416,4452,4458,4489,4728,4890~\AA, which originate in the outer stellar wind. We found  v$_{\textnormal{\scriptsize{sys}}} =$~192$\pm$3~km~s$^{-1}$, in good agreement with the literature values v$_{\textnormal{\scriptsize{sys}}} = 195$~km~s$^{-1}$ \citep{1974MNRAS.168..221T} and v$_{\textnormal{\scriptsize{sys}}} = 193$~km~s$^{-1}$ \citep{1981A&A...103...94W}, which were also determined from [\ion{Fe}{II}] lines. All radial velocities stated below are relative to the systemic velocity.

\section{Results}
\label{results}

The 2012 X-shooter spectra confirm the dramatic spectral changes reported in \citet{2012CBET.3192....1G}. However, the peak visual magnitude appears to have been lower (see http://www.aavso.org) as stated by Gamen et al.\ and we do not confirm that R71 currently resembles an early G-supergiant. Figure \ref{figure:overview} compares a UVES spectrum obtained in 2002 July with an X-shooter spectrum obtained in 2012 November.  Most emission lines have disappeared and the spectrum is dominated by strong neutral and singly ionized metal absorption lines  (e.g., \ion{Fe}{I}, \ion{Fe}{II}, \ion{Si}{I}, \ion{Si}{II}, \ion{S}{I}, \ion{N}{I}, and \ion{O}{I}) indicative of a cold dense pseudo-photosphere, which develops during an LBV outburst \citep{1985A&A...153..168L,1986IAUS..116..139A,1987ApJ...317..760D}. 

We analyzed several spectral features between 2002--2012. The UVES spectra in 2002 were obtained during the maximum of a microvariation ($m_{\textnormal{\scriptsize{V}}}\sim10.7$~mag compared to $m_{\textnormal{\scriptsize{V}}}\sim10.9$~mag in quiescence, see Figure 1 in \citealt{2010AJ....140...14S}). Small variations in the line profiles occur during 2002 but the spectra nevertheless closely resemble R71's appearance in quiescent phase.\footnote{No significant line variations hinting at a  potential companion is observed during the 3 months of UVES coverage. Comparison of the UVES spectra to the FEROS and X-shooter spectra with respect to line variations due to binarity is inhibited because of the spectral changes inflicted by the eruption.} At the time the 2005 FEROS spectrum was obtained, the current eruption had already commenced ($m_{\textnormal{\scriptsize{V}}}\sim10.7$~mag). 
The 2002 July and 2005 December spectra show no major differences. 
The 2007 FEROS spectrum appears to be a snapshot of the transitionary phase between the quiescent and the eruptive state with $m_{\textnormal{\scriptsize{V}}}\sim10.3$~mag. The 2011 FEROS and 2012 X-shooter spectra show R71 in eruption when the visual magnitude reached up to $m_{\textnormal{\scriptsize{V}}}\sim8.7$~mag (http://www.aavso.org, cf. \citealt{2012CBET.3192....1G} who observed a maximum brightness of $m_{\textnormal{\scriptsize{V}}}\sim8.3$~mag). 

\begin{figure}
\centering
\resizebox{\hsize}{!}{\includegraphics{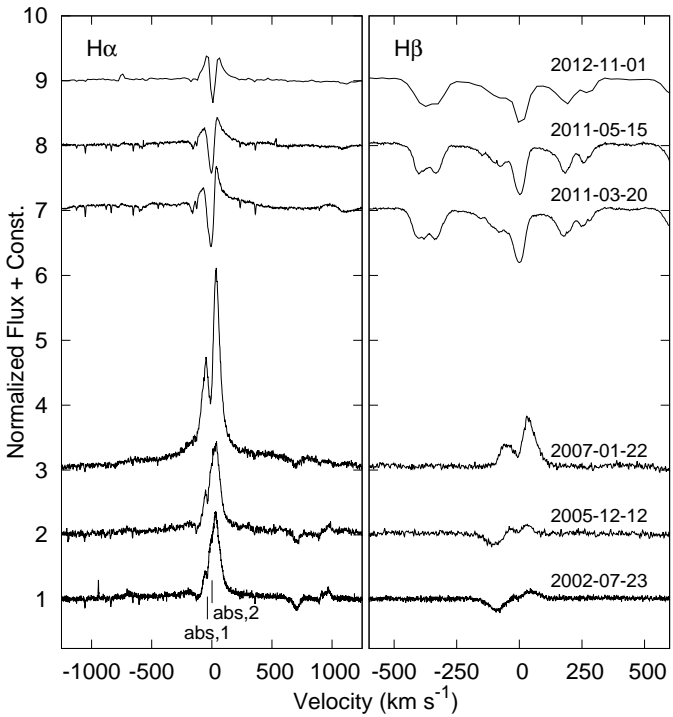}}
     \caption{H$\alpha$ and H$\beta$ in R71 from 2002--2012. The H$\alpha$ P Cyg profile changed to an inverse double-peaked symmetric profile during the eruptive state. The absorption component close to system velocity now dominates the profile.  H$\beta$ behaves similarly to H$\alpha$. In the latest spectrum from 2012 November, the \ion{Fe}{I} $\lambda$6546 emission line becomes apparent towards the blue of H$\alpha$. The strong absorption features to the red and blue of H$\beta$ in 2011--2012 are blends of \ion{Fe}{I}, \ion{Ni}{I}, \ion{Ti}{I}, \ion{Ti}{II} absorption lines. (The continuum is normalized to unity, velocities are in R71's restframe.)}
     \label{figure:Halpha}
\end{figure}

Figure \ref{figure:Halpha} shows a time series of H$\alpha$ and H$\beta$ from 2002--2012.
H$\alpha$ changed from a P Cyg profile with broad emission wings extending up to v$_{\textnormal{\scriptsize{em}}}\sim\pm850$~km~s$^{-1}$ in 2002--2005 to a prominent double-peaked symmetric profile in 2011/2012. 
The maximum depth of the P Cyg absorption in 2002 July is at v$_{\textnormal{\scriptsize{abs}}} = -$118$\pm$12~km~s$^{-1}$  and the blue edge is at v$_{\textnormal{\scriptsize{edge}}} = -$186$\pm$5~km~s$^{-1}$. We use the blue edge of the H$\alpha$ P Cyg absorption as a proxy for the terminal velocity but since the absorption is not well-defined and filled in with extra emission this measurement gives only a lower limit. The profile shows two additional absorption features at v$_{\textnormal{\scriptsize{abs,1}}} = -$37$\pm$2~km~s$^{-1}$ and  at v$_{\textnormal{\scriptsize{abs,2}}} =$~2$\pm$6~km~s$^{-1}$. The former may be due to an expanding shell, while the latter originates most likely from the photosphere.
In spectra obtained between 2002 August and October we find  v$_{\textnormal{\scriptsize{edge}}} = -$131$\pm$20~km~s$^{-1}$.
In 2007 extra emission compared to the continuum is observed. The P Cyg absorption has disappeared as has the absorption component at v$_{\textnormal{\scriptsize{abs,1}}}  \sim -37$~km~s$^{-1}$. The weak absorption component at system velocity strengthened, most likely due to the formation of a cooler pseudo-photosphere. 
In 2011/2012 this absorption component dominates the H$\alpha$ profile. It has a full width at half maximum (FWHM) of  44$\pm$1~km~s$^{-1}$. There was no indication of this absorption component during the previous outburst but this absorption feature was first observed in the 1990s.  Weak broad H$\alpha$ emission is still present and a weak absorption component can be observed with a  blue edge at v$_{\textnormal{\scriptsize{edge}}} = -$199$\pm$9~km~s$^{-1}$. 

We found comparable values for the blue edge of the H$\alpha$ absorption during quiescence and eruption. However, because the H$\alpha$ absorption depends on the density and ionization structure of the wind, and the star is currently hidden beneath a pseudo-photosphere, we cannot state for certain whether the wind velocity stayed unchanged during the current outburst. Our values of v$_{\textnormal{\scriptsize{edge}}} \sim -131$~km~s$^{-1}$ to $-$199~km~s$^{-1}$ are comparable to the value v$_{\textnormal{\scriptsize{edge}}} = -158$~km~s$^{-1}$ in 1984 found by \citet{1986A&A...158..371S}. \citet{1981A&A...103...94W}  found a maximum outflow velocity of $\sim$127~km~s$^{-1}$ in UV \ion{Fe}{II} lines.

H$\beta$ behaves similarly to H$\alpha$ but does not show prominent line emission in 2011/2012. 
Higher Balmer lines are always in absorption.  Paschen lines and Br$\gamma$, which were likely in emission during quiescent state, are in absorption in 2012.\footnote{No archival near-infrared spectra of R71 are available and since this paper focuses on the comparison between R71's current eruption with its quiescent phase and its previous outburst we will only mention a few near-infrared features. R71's near-infrared spectrum during the current eruption will be discussed in more detail in a forthcoming paper.} The absence of hydrogen emission (apart from H$\alpha$ emission) indicates that R71's pseudo-photosphere is much cooler and denser during the current eruption than during its 1970s outburst when H$\alpha$ to H$\delta$ showed strong P Cyg profiles \citep{1981A&A...103...94W}.

\begin{figure}
\centering
\resizebox{\hsize}{!}{\includegraphics{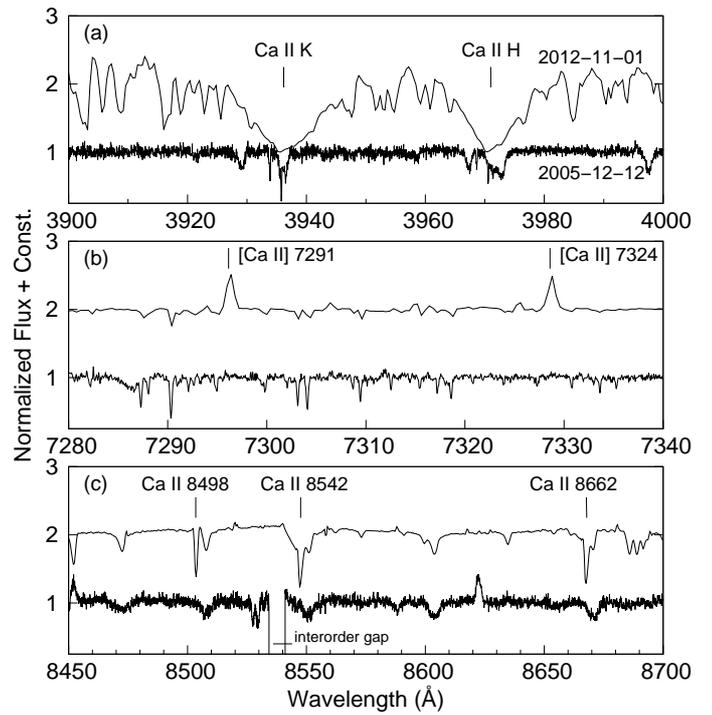}}
     \caption{\ion{Ca}{II} lines in 2005 and 2012; (a) the \ion{Ca}{II} H and K lines show very broad and deep absorption, (b) the [\ion{Ca}{II}] $\lambda\lambda$7291,7324 lines are in emission, and (c) the \ion{Ca}{II} near-infrared lines at $\lambda\lambda$8498,8542,8662~\AA\ are in absorption. The Paschen lines $\lambda\lambda$8502,8545,8665~\AA\ can be seen in absorption just to the right of the  \ion{Ca}{II} triplet lines.  (The continuum is normalized to unity.)}
     \label{figure:CaII}
\end{figure}

Figure \ref{figure:CaII} compares \ion{Ca}{II} lines in 2005 and in 2012. These lines provide information on the (changing) conditions of the pseudo-photosphere and the circumstellar material. \ion{Ca}{II} has a low ionization potential and the broad and very deep \ion{Ca}{II} H and K absorption lines observed in 2011 and 2012 are therefore an indicator of the low temperatures in R71's pseudo-photosphere.  The forbidden [\ion{Ca}{II}] $\lambda\lambda$7291,7324 lines are in emission and the near-infrared triplet \ion{Ca}{II} $\lambda\lambda$8498,8542,8662 is in absorption.
The \ion{Ca}{II} near-infrared triplet lines are usually observed in emission in warm hypergiants. They are formed by radiative de-excitation from the upper level of the \ion{Ca}{II} H and K absorption, which leaves the Ca$^+$ ions in the upper level for the [\ion{Ca}{II}] $\lambda\lambda$7291,7324 lines.\footnote{Alternatively, \citet{2009ApJ...705.1425P} and \citet{2009ApJ...697L..49S} have suggested that the [\ion{Ca}{II}] $\lambda\lambda$7291,7324  emission lines may be due to dust destruction. However, it is unclear what would cause the destruction of dust during this eruption.} We investigated the line variations in our 2012 spectra in order to explain the puzzling combination of \ion{Ca}{II} near-infrared triplet absorption and  [\ion{Ca}{II}] $\lambda\lambda$7291,7324  emission observed in R71. 
We find that the \ion{Ca}{II} near-infrared triplet lines show the same residuals as the H$\alpha$ emission, while pure photospheric lines vary little. The \ion{Ca}{II} triplet is likely dominated by absorption but filled in by an emission component from the circumstellar material that may be sufficient to populate the upper level of the [\ion{Ca}{II}] $\lambda\lambda$7291,7324 lines.
%But since the strong \ion{Ca}{II} H and K absorption lines populate the upper level, there is no need to invoke this process.

Other spectral lines confirm the very low temperature of R71's pseudo-photosphere.  In the previous outburst almost all lines (in particular the Balmer lines, \ion{Fe}{II}, and \ion{Ti}{II}) showed classical P Cyg profiles with a deep absorption component \citep{1981A&A...103...94W}, which is not the case for the current eruption.  [\ion{Fe}{II}] emission lines have disappeared. \ion{Mg}{II} $\lambda$4481 absorption, indicative of low temperatures, is strong.  \ion{He}{I} $\lambda\lambda$4388,4471,5876,6678 absorption at system velocity in pre-eruption spectra have disappeared in 2011/2012. We have no comparison spectra for the near-infrared \ion{He}{I} $\lambda$10830 line during the quiescent state and thus cannot conclusively identify the absorption feature and the small emission observed in 2012 in the concerning wavelength region as \ion{He}{I}. More likely no \ion{He}{I} absorption and emission lines are present at the current time. During the 1970s outburst the spectrum showed \ion{He}{I} absorption \citep{1974MNRAS.168..221T}, again indicating lower temperatures during the current eruption.
The \ion{Na}{I} $\lambda\lambda$5890,5896 lines show three absorption components: 1.) the Galactic interstellar absorption with v~$ = -$177$\pm$1~km~s$^{-1}$  relative to R71, 2.) an absorption component with v~$=$~13$\pm$1~km~s$^{-1}$, which likely is the interstellar LMC feature, and 3.) a component with v~$= -$29$\pm$1~km~s$^{-1}$, which is conceivably formed in the cool pseudo-photosphere and the circumstellar material based on its line strength variations. The velocity of this component stayed constant compared to pre-outburst, but the line became stronger.

Between 2012 August and November \ion{Fe}{I} emission lines appeared redwards of their absorption components (producing a P Cyg-like profile). In Figure \ref{figure:Halpha}, e.g., the \ion{Fe}{I} $\lambda$6546 line can be seen in the latest spectrum (at about $-$750~km~s$^{-1}$ from H$\alpha$).  According to \citet{2011A&A...526A.116K} the observed \ion{Fe}{I} $\lambda$6430 to \ion{Fe}{I} $\lambda$6421  ratio of $\sim$2, i.e., different from about unity, would indicate a non-thermal excitation mechanism.
Other emission lines characteristic of a very low-excitation ``nebula'' (excitation energy on the order of $4$~eV and less) appeared and became stronger.
 We tentatively identify narrow emission from, e.g., [\ion{O}{I}] $\lambda\lambda$6300,6364,6392, [\ion{Ni}{I}] $\lambda$7394, and [\ion{S}{I}] $\lambda\lambda$7725,10821. These lines indicate the presence of a  neutral nebula, likely ejected during the current eruption and possibly the rarefied region above the pseudo-photosphere. The \ion{Fe}{I} emission lines, which appeared in late 2012, may originate from the same region. Classical nebular lines from an ionized nebula such as [\ion{N}{II}] $\lambda\lambda$6548,6583, which were observed in R71's spectrum in 1984 \citep{1986A&A...158..371S}, are not present. The expansion velocity of the nebula is on the order of a few  km~s$^{-1}$ and thus small compared to the stellar wind velocity. The eruption started about 6 years ago and, if we assume an expansion velocity of v$_{\textnormal{\scriptsize{eruption}}} \sim10$~km~s$^{-1}$, the ejected material is now reaching a distance of about 13~AU or $3\,000~R_{\odot}$ from the star. At a distance of the LMC, this corresponds to about 0.3~mas.
 
Photometrically and spectroscopically, R71's current eruption is very different compared to its previous outburst. Its visual lightcurve reached a peak magnitude in 2012 that was about 1~mag brighter than during the 1970s outburst and R71's current spectrum implies a much cooler pseudo-photosphere.

\begin{table*}
\caption{Parameters of R71.} % title of Table
\label{table:1} % is used to refer this table in the text
\centering % used for centering table
\begin{tabular}{l l c c c c c c c } % centered columns (4 columns)
\hline\hline % inserts double horizontal lines
State & Source  & $m_{\textnormal{\scriptsize{V}}}$   & $DM_{\textnormal{\scriptsize{LMC}}}$& $A_{\textnormal{\scriptsize{V}}}^{a}$ &  $T_{\textnormal{\scriptsize{eff}}}$ & $BC$ & log ($L / L_{\odot}$)$^{b}$ &  $M_{\textnormal{\scriptsize{bol}}}^{b}$  \\ % table heading
 &   & (mag) & (mag) & (mag) & (K) &  (mag) &  & (mag) \\ % table heading
\hline % inserts single horizontal line
Quiescence &  Wolf et al.\ (1981) & 10.88 & 18.50 &  0.15 & $13\,600$ & -0.76  & 5.30 & -8.53  \\
 & van Genderen (1982) & 10.89  & 18.60 &  0.15 & $14\,000$ & -0.95   & 5.37 & -8.71 \\ 
 & van Genderen (1988)  & 10.87 & 18.60 & 0.37 & $14\,700$ & -1.10    & 5.53 & -9.10 \\ 
  & Lennon et al.\ (1993) & 10.88 & 18.45 & 0.63& $17\,250$   & -1.69  & 5.86 & -9.94 \\ 
  &  & & & & & &\\
Outburst 1970--1977 & van Genderen (1982) & 9.79 &  18.60 &  0.15& $11\,000$   & -0.38  & 5.58 & -9.24 \\
& van Genderen (1988) & 9.77 &  18.60 &  0.37& $8\,150$    & -0.10  & 5.57  & -9.20 \\
 &  & & & & & &\\
Outburst 2012 & this paper & 8.7$^{c}$ &  18.50$^{d}$ & 0.15$^{e}$ &  6\,650 & 0.11   & 5.82 & -9.84  \\
 &  & 8.7$^{c}$& 18.50$^{d}$   & 0.37$^{f}$& 6\,650  & 0.11 & 5.91 &  -10.06 \\
 & & 8.7$^{c}$ & 18.50$^{d}$ & 0.63$^{g}$  & 6\,650 & 0.11 &  6.02 & -10.32 \\
\hline %inserts single line
\end{tabular}
\tablebib{
(a) In the cases were only $E(B$--$V)$ was given in the literature, we adopted the standard extinction law with $R_{\textnormal{\scriptsize{V}}} = A_{\textnormal{\scriptsize{V}}}/E(B$--$V)= 3.1$ \citep{1983MNRAS.203..301H}; (b) values were calculated using $DM_{\textnormal{\scriptsize{LMC}}}=18.5$~mag; (c) http://www.aavso.org, but see \citet{2012CBET.3192....1G}; (d) e.g., \citet{2001ApJ...553...47F};
%,2006ApJ...652.1133M,2007AJ....133.1810B,2008glv..book..317N,2011A&A...534A..95S,2012ApJ...747...50N,2012ApJ...758...24F,2012arXiv1212.4376I}; 
(e) \citet{1981A&A...103...94W}; (f) \citet{1988A&AS...74..453V}; (g) \citet{1993SSRv...66..207L}.
}
\end{table*}

\section{Discussion}
\label{discussion}

R71 was well-observed over the last decades covering its quiescent phases and its 1970--1977 classical LBV outburst. This gives us a unique opportunity to compare the current eruption to its previous behavior. Table \ref{table:1} gives an overview of R71's parameters found for its quiescent state and its 1970s outburst by different authors, and parameters estimated in this paper for its current eruption.

There is ambiguity in the literature regarding the classification of R71 as a classical LBV \citep{1993SSRv...66..207L} or as a less luminous LBV \citep{1981A&A...103...94W}. Reasons are the uncertainty in the extinction towards R71 and the bolometric correction, see Table \ref{table:1}. To interpret R71's behavior it is important to know to which of the two groups it belongs because they differ in mass and evolutionary path \citep{1994PASP..106.1025H}. Classical LBVs have $M_{\textnormal{\scriptsize{bol}}} < -9.6$~mag and have very likely not been RSGs. Less luminous LBVs have $M_{\textnormal{\scriptsize{bol}}} = -8$~mag to $-9$~mag, lower temperatures, smaller amplitudes of their outbursts, and lower mass loss rates. They have probably been RSGs. 

Support for the classification of R71 as a less luminous LBV comes from the fact that its visual light and temperature variation during the 1970s outburst showed much smaller variations than is observed for the more luminous LBVs  \citep{1974MNRAS.168..221T,1981A&A...103...94W}. Further evidence in favor of a RSG phase  comes from infrared studies. A dust shell around R71 may have been produced in a short RSG phase \citep{1999A&A...341L..67V,2010AJ....139...68V,2010A&A...518L.142B}. If the period-luminosity relation for LBVs found by  \citet{1995ApJ...451L..61S} holds for R71, then the timescale of about 40 years between the current eruption and its 1970s outburst implies $M_{\textnormal{\scriptsize{bol}}} \sim -9$~mag (only 20 years between outbursts would have been expected in case of $M_{\textnormal{\scriptsize{bol}}} \sim -10$~mag). 
\citet{1993SSRv...66..207L}, however, argued that the extinction towards R71 is anomalous and greater than previously thought and thus found a much higher luminosity and temperature, more consistent with a spectral type of B2.5 Ieq. R71 enters the region of classical LBVs in the HR-diagram, if we adopt the extinction value found by \citet{1993SSRv...66..207L}. Its 1970s outburst may have been unusually weak compared to other classical LBV outbursts.

R71 does not appear to lie in a region with anomalous extinction, see, e.g., Figure 11 in \citet{2007ApJ...662..969I} and Figure 5 in \citet{2008A&A...484..205D}. The empirical relation between sodium absorption and dust extinction found by \citet{2012MNRAS.426.1465P} and $R_{\textnormal{\scriptsize{V}}} = 3.1$ implies an interstellar extinction of $A_{\textnormal{\scriptsize{V}}} =$~0.09$\pm$0.03~mag towards R71. This value is likely too small. \citet{2010A&A...518L.142B}, e.g., assumed $A_{\textnormal{\scriptsize{V}}} = 0.4$~mag based on the extinction map by \citet{1998ApJ...500..525S}. A potential larger extinction towards R71 would likely be of circumstellar nature and may be caused by dust formed in carbon deficient ejecta \citep{1993SSRv...66..207L}. 
An upper limit to the extinction towards R71 of $A_{\textnormal{\scriptsize{V,2012}}} <$~0.7$\pm$0.4~mag can be inferred from the absence of the $\lambda$10780 absorption line for which a relation between the line equivalent width and $E(B$--$V)$ was found by \citet{2007A&A...465..993G}. 
While this upper limit does not discriminate against any of the extinction values found in the literature, it implies that the current extinction value is certainly not much higher than the pre-eruption value. The use of other diffuse interstellar medium lines and their correlation with extinction is inhibited because line identification is made impossible by many small emission and absorption lines even in R71's quiescent state. 
R71 has likely an extinction value of $ 0.1 < A_{\textnormal{\scriptsize{V}}} < 1.1$~mag and thus lies close to the boundary between the less luminous and the classical LBVs during its quiescent state and a previous short RSG phase cannot entirely be ruled out.

The observed spectral changes during R71's current eruption are expected in an outburst that produces an optically thick wind \citep{1994PASP..106.1025H} but it is not well-established if LBVs do form pseudo-photospheres. \citet{1989ApJ...346..919L} and \citet{1996A&A...306..501D} found based on NLTE modeling that the observed variability of the photospheric radius and effective temperature cannot be due to the formation of a pseudo-photosphere but must be induced by  a sub-photospheric instability. \citet{2004ApJ...615..475S}, on the other hand, argued that objects close to the bi-stability jump and close to the Eddington limit could form pseudo-photospheres. In our discussion we adopt the pseudo-photosphere hypothesis, i.e., that the apparent changes in R71's spectrum and photometry are caused in the outermost layers. However, we are aware that the discussion is still open on this issue.
 
Estimations of R71's current effective temperature based on intrinsic colors are likely to fail because LBVs and R71 in particular show generally a UV excess \citep{1980A&A....88...15W,1993SSRv...66..207L} and the reddening towards R71 is disputed.
For example, the relation between $T_{\textnormal{\scriptsize{eff}}}$ and $(B$--$V)_{\textnormal{\scriptsize{0}}}$ found by \citet{2010ApJ...719.1784N} results in $T_{\textnormal{\scriptsize{eff}}} =$~5\,600--6\,000~K if we adopt $(B$--$V) = 0.6$~mag (http://www.aavso.org) and allow for different extinction values. However, spectral features imply a higher effective temperature.
To determine R71's apparent temperature we used the line-depth ratios described in \citet{2000A&A...358..587K}. We determined a current effective temperature of $T_{\textnormal{\scriptsize{eff}}} =$~6\,700$\pm$400~K. 
This corresponds to an apparent spectral type of F5 to F8 and a bolometric correction of $BC=0.08$--$0.13$~mag \citep{1984ApJ...284..565H}.

We also compared the 2012 X-shooter spectra to Kurucz model atmospheres \citep{1997A&A...318..841C}. 
We varied the metallicity from [Fe/H] $= 0$ to [Fe/H] $= -0.5$ (close to the literature value [Fe/H]$_{\textnormal{\scriptsize{LMC}}}  = -0.34$, see \citealt{1998AJ....115..605L}), the effective surface gravity $\log g_{\textnormal{\scriptsize{eff}}}  = 0.0$ to $\log g_{\textnormal{\scriptsize{eff}}}  = 2.0$, and the extinction $A_{\textnormal{\scriptsize{V}}} = 0.15$ to $A_{\textnormal{\scriptsize{V}}} = 0.63$~mag. 
Two immediate problems arise when comparing the X-shooter spectra to model atmospheres. First, LBVs have weaker Balmer jumps than normal stars. \citet{2005ChJAA...5..245G} suggested that this may be due to a decrease of neutral hydrogen because of collisions in the dense wind, due to extra emission from the recombination of hydrogen, or due to the more extended atmospheres in which the velocity field changes with the radius and the lines of the Lyman series could be spread out. \citet{2007ApJ...659.1563G} showed that the continuum energy distribution around the Balmer jump is sensitive to the wind velocity law. Second, the observed spectral energy distribution of LBVs in outburst may be a combination of two sources (the central star and the envelope). \citet{2005ChJAA...5..245G} proposed that LBVs form a non-homogeneous envelope during outburst, which is not optically thick in all directions. The observed spectral energy distribution comes from the central star (mostly in the UV) as well as the optically thick part of the envelope (mostly in the optical wavelength regions). 
When comparing model atmospheres to the X-shooter spectra we thus focus on the continuum emission redwards of the Balmer jump.

Models with higher surface gravity $\log g$ have weaker Balmer and Paschen jumps. Because of the above described issues, we make an apriori assumption about  $\log g$.
\citet{1993SSRv...66..207L} found $\log g=1.8$ for R71 in quiescent state. The effective surface gravity in 2012 is likely  $\log g_{\textnormal{\scriptsize{eff,2012}}} \lesssim 0.5$ based on the fact that  R71's effective radius has increased by up to a factor of 5, see below, and that low effective gravities of $\log g_{\textnormal{\scriptsize{eff}}} = 0.5$--$1.0$ were found for several LBVs in outburst. For example, \citet{1991A&A...247..383S} estimated $\log g_{\textnormal{\scriptsize{eff}}} = 0.55$ for HD 160529 and \citet{1993A&A...280..508S} found $\log g_{\textnormal{\scriptsize{eff}}} = 0.75$ for R40. Both objects are similar to R71. \citet{2005ChJAA...5..245G} determined $\log g_{\textnormal{\scriptsize{eff}}} = 1$ for R127 and $\log g_{\textnormal{\scriptsize{eff}}} = 0.5$ for R110.
We thus adopt $\log g_{\textnormal{\scriptsize{eff,2012}}} = 0.5$ for R71, also on the grounds that for lower surface gravities the models become unstable against radiation pressure.

\begin{figure}
\centering
\resizebox{\hsize}{!}{\includegraphics{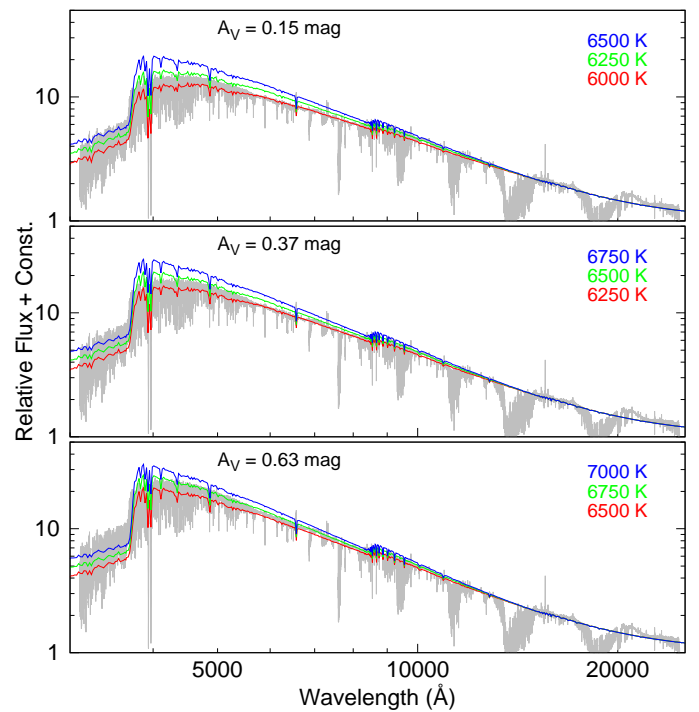}}
     \caption{Comparison of de-reddened X-shooter spectra from 2012 November and Kurucz model atmospheres (with fixed parameters [Fe/H] = $-$0.5, $\log g = 0.5$). The continuum energy distribution is well fitted for the extinction values $A_{\textnormal{\scriptsize{V}}}$ found in the literature. Higher extinction values require models with slightly higher effective temperatures.}
     \label{figure:sedfit}
\end{figure}

Figure \ref{figure:sedfit} shows de-reddened ($A_{\textnormal{\scriptsize{V}}} = 0.15, 0.37, 0.63$~mag) 2012 November X-shooter spectra and model atmospheres of different effective temperatures.  The analytic formula for the mean extinction law by \citet{1989ApJ...345..245C}  and $R_{\textnormal{\scriptsize{V}}}=3.1$ \citep{1983MNRAS.203..301H} were adopted to de-redden the X-shooter spectra. In each case the best fit leads to a too high Balmer discontinuity, but see above. Unfortunately, we cannot discriminate between the different values of $A_{\textnormal{\scriptsize{V}}}$ used in the literature. For higher extinction values the X-shooter spectra are well fitted by model atmospheres with higher effective temperatures.  
We find that model atmospheres with  $T_{\textnormal{\scriptsize{eff}}} =$~6\,600$\pm$300~K  fit best the spectral energy distribution of our X-shooter spectra, if we take into account the range of extinction values $A_{\textnormal{\scriptsize{V}}} = 0.15$--$0.63$~mag. The derived effective temperature from model atmospheres is very similar to the value from the line ratio measurements discussed above.  
R71's pseudo-photosphere has certainly reached much lower temperatures than during its 1970s outburst when the apparent temperature was not cooler than $T_{\textnormal{\scriptsize{eff}}} \sim 8\,150$~K. However, we do not confirm the result by \citet{2012CBET.3192....1G} that R71 has currently an early G-supergiant spectrum.

While there is uncertainty of R71's parameters during its quiescent state because of uncertainties in the extinction $A_{\textnormal{\scriptsize{V}}}$ and the bolometric correction $BC$ (see Table \ref{table:1}), its current temperature implies a small value for $BC$. We adopt a distance modulus of  $DM_{\textnormal{\scriptsize{LMC}}} = 18.5$~mag (e.g., \citealt{2001ApJ...553...47F}), a visual extinction of $A_{\textnormal{\scriptsize{V}}} = 0.15$--$0.63$~mag, and a bolometric correction of $BC = 0.11$~mag, and find that R71's current visual brightness of $m_{\textnormal{\scriptsize{V,2012}}} = 8.7$~mag results in $M_{\textnormal{\scriptsize{bol,2012}}}\sim-9.8$~mag to $-10.3$~mag. R71's luminosity of $\log L_{\textnormal{\scriptsize{R71,2012}}} / L_{\odot}$ $= 5.8$--$6.0$ is about its classical Eddington luminosity ($\log L_{\textnormal{\scriptsize{R71,Edd}}}/L_{\odot} \sim 5.8$--$6.1$). The pre-eruption bolometric luminosity was $M_{\textnormal{\scriptsize{bol,quiescence}}}\sim-8.5$~mag to $-9.9$~mag if we adopt $DM_{\textnormal{\scriptsize{LMC}}} = 18.5$~mag instead of the varying distance moduli used in the literature.\footnote{The range of the pre-eruption bolometric luminosity is larger because of the greater uncertainty in the bolometric correction $BC$.} 
The bolometric luminosity thus increased by 0.4--1.3~mag.

The very low effective temperature of R71's pseudo-photosphere during the current eruption below $7\,500$~K implies a very high wind density \citep{1987ApJ...317..760D}. 
LBVs in quiescence have mass loss rates comparable to normal supergiants of the same temperature and luminosity ($10^{-7}$~$M_{\odot}$ yr$^{-1}$ to $10^{-5}$ ~$M_{\odot}$ yr$^{-1}$), which increase 10--100 times during an outburst.
R71's mass loss rate during its quiescent state was 3$\times$10$^{-7}$~$M_{\odot}$~yr$^{-1}$ and during its 1970s outburst 5$\times$10$^{-5}$  $M_{\odot}$ yr$^{-1}$ \citep{1981A&A...103...94W}.  
We use Figure 1 and Equation 4 in \citet{1987ApJ...317..760D} for a rough estimate of R71's current mass loss rate and find $\dot M_{\textnormal{\scriptsize{R71,2012}}} \sim 7\times$10$^{-5}$~$M_{\odot}$~yr$^{-1}$ to 7$\times$10$^{-3}$~$M_{\odot}$~yr$^{-1}$. Equation 1 in \citet{1980A&A....88...15W} results in a current mass loss rate of $\dot M_{\textnormal{\scriptsize{R71,2012}}} \sim$~2$\times$10$^{-4}$~$M_{\odot}$~yr$^{-1}$.
R71 thus increased its mass loss rate by a factor of about $1\,000$. 
The total mass of ejected material during the last 6 years may therefore be on the order of 10$^{-3}$~$M_{\odot}$. 
We find no evidence of an explosion. The spectra show no signatures of fast moving material and the terminal velocity  is comparable to R71's quiescent state. Also, during its 1970s outburst the terminal velocity was comparable \citep{1981A&A...103...94W}. This indicates that the change in mass loss rate is caused by a tremendously increased wind density only.

The effective radius increased by a factor of about 5 from 81--95~$R_{\odot}$ \citep{1981A&A...103...94W,1993SSRv...66..207L} during the quiescent phase and is now on the order of $500~R_{\odot}$. This radius estimation is based on the apparent temperature and luminosity during the eruption and may not be very meaningful because the opacity in the pseudo-photosphere is mainly due to scattering \citep{1987ApJ...317..760D}. 
%The different values used for  $DM$, $T_{\textnormal{\scriptsize{eff}}}$, and $BC$ by the authors contribute as much as the different extinction values to the discrepancies in R71's total luminosity found in the literature. 

\begin{figure}
\centering
\resizebox{\hsize}{!}{\includegraphics{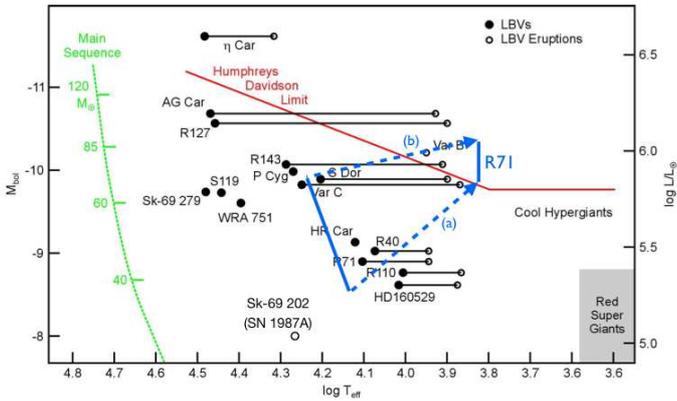}}
     \caption{Schematic upper HR-diagram (modified from \citealt{1994PASP..106.1025H}). The red solid curve is the upper luminosity boundary. Black solid lines represent LBV transitions. The positions of R71 during its quiescent state and its current eruption are indicated with blue solid curves, see text (R71's quiescent and 1970s outburst locations placed by \citealt{1994PASP..106.1025H} are also shown). The dashed blue curves show the transitions for the lowest and highest extinction values found in the literature; (a) $A_{\textnormal{\scriptsize{V}}} = 0.15$~mag and (b) $A_{\textnormal{\scriptsize{V}}} = 0.63$~mag.}\label{figure:HRdiagram}
\end{figure}

Figure \ref{figure:HRdiagram} shows the location of R71 in quiescence and during its current eruptive state in the HR-diagram. The solid blue lines indicate the range of bolometric luminosities obtained for each state, if we adopt $DM_{\textnormal{\scriptsize{LMC}}} = 18.5$~mag and the different values for  $A_{\textnormal{\scriptsize{V}}}$ and $BC$ found in the literature, see Table \ref{table:1}. The dashed blue lines  indicate the transitions for the two extreme cases. The transition labelled `(a)' is the transition for the lowest extinction value, $A_{\textnormal{\scriptsize{V}}}=0.15$~mag, and the transition labelled `(b)' is the transition for the highest extinction value $A_{\textnormal{\scriptsize{V}}}=0.63$~mag. R71 could have potentially increased its bolometric luminosity by up to 1.3~mag.

R71's visual magnitude variation on the order of $\Delta m_{\textnormal{\scriptsize{V}}} \sim 2$~mag  implies that the star is experiencing a classical LBV outburst \citep{1994PASP..106.1025H}. However, the effective temperature appears to be unusually low and the mass loss rate unusually high. Also, the common notion is that during classical LBV outbursts the bolometric luminosity stays constant, which is most likely not the case for R71's current eruption. With a growing number of observed LBV outbursts and with the realization that outbursts of the same object can differ considerably this commonly adopted classification scheme may have to be reconsidered.

\section{Conclusion}
\label{conclusion}

R71's current eruption differs from its 1970s classical LBV outburst in several aspects and is challenging our view on LBVs. Both the visual lightcurve and the spectra show many differences compared to its previous outburst.
The visual light increased by $\Delta m_{\textnormal{\scriptsize{V}}} \gtrsim 2$~mag over the last seven years and the spectrum indicates an unusually low temperature of R71's pseudo-photosphere, mimicking the photosphere of a late-F supergiant. Balmer and \ion{Fe}{II} P Cyg profiles, normally observed during LBV outbursts and present during R71's previous 1970s outburst, are absent. Low-excitation forbidden emission lines and \ion{Fe}{I} P Cyg-like profiles appear in late 2012.

We found an apparent temperature of R71 in 2012 of $T_{\textnormal{\scriptsize{eff,2012}}}\sim$~6\,650~K. This is a decrease by more than 7\,000~K compared to its quiescent state and more than 1\,500~K cooler than during its 1970s outburst.
The bolometric luminosity increased by 0.4--1.3~mag to $M_{\textnormal{\scriptsize{bol,2012}}}\sim-9.8$~mag to $-10.3$~mag and R71 may have moved beyond the Humphreys-Davidson limit. R71's apparent radius increased by a factor of 5 to about $500~R_{\odot}$. 

We estimated that R71 has a current mass loss rate on the order of $\dot M_{\textnormal{\scriptsize{R71,2012}}} \sim$~5$\times$10$^{-4}
$~$M_{\odot}$~yr$^{-1}$,  a factor of about $1\,000$ higher than during its quiescent state and at least a factor of 10 larger than during its 1970s outburst. 
The wind velocity of R71 is on the order of 100--200~km~s$^{-1}$, similar to that of other LBVs while the pseudo-photosphere has a velocity on the order of a few km~s$^{-1}$. We find no spectral signatures of fast moving material, which may indicate an explosion. The large increase in mass loss rate during the eruption is therefore likely due to an increased wind density only, thus constraining instability mechanisms. 

LBVs have been considered as extragalactic distance indicators because they belong to the most luminous stars (e.g., \citealt{1968ApJ...151..825T,1983AJ.....88.1569S,1984AJ.....89..630S}). \citet{1989A&A...217...87W} found a relation between the luminosity of LBVs and their outburst amplitudes. However, with an increasing number of observed LBV outbursts it becomes clear that they can differ considerably -- even outbursts of the same object. S Dor, e.g., displayed an early F-type supergiant spectrum in 1999, which had never been reported before, while its visual luminosity only increased by 0.3~mag \citep{2000PASP..112..144M}.  The current eruption of R71 has a much larger amplitude than its previous outburst in the 1970s. Thus LBVs appear to be unreliable as distance indicators. 

R71 is experiencing a much more powerful LBV eruption than during its 1970s outburst. We are in the unique position to witness such a rare event as it unfolds. 
Our X-shooter monitoring program on R71 will secure key data for our understanding of LBV outbursts and the properties of stars near the upper luminosity limit. 
Observations in the infrared after the eruption has subsided are highly desirable because they will provide valuable information about the total amount of material expelled.

\begin{acknowledgements} We thank the Paranal Observatory for conducting the observations and ESO for the acceptance of our DDT program. 
We also thank Roberta M. Humphreys and Willem-Jan de Wit for valuable discussions. 
\end{acknowledgements}

\bibliographystyle{aa}

\end{document}